\documentclass[aps,pre,onecolumn,superscriptaddress,amsmath,amssymb]{revtex4-2} 

\usepackage{amssymb,amsmath,amsbsy}
\usepackage{upgreek}
\usepackage{cancel}
\usepackage{mathdots}
\usepackage{stackrel}
\usepackage{enumerate}
\usepackage{mathrsfs}
\usepackage{graphicx}
\usepackage{float}
\usepackage{mathrsfs}
\usepackage{listings}
\usepackage{hyperref}
\usepackage{pgf}
\usepackage{tikz}
\usetikzlibrary{arrows.meta, positioning, calc, 3d, decorations.pathmorphing}
\usetikzlibrary{decorations.pathmorphing, arrows.meta, positioning}
\usetikzlibrary{external}
\usepackage{polynom}
\usepackage{soul}

\newcommand{\sgn}{\operatorname{sgn}}

\begin{document}

\newcommand\chau{\affiliation{ 
 Charles University,  
 Faculty of Mathematics and Physics, 
 Department of Macromolecular Physics, 
 V Hole{\v s}ovi{\v c}k{\' a}ch 2, 
 CZ-180~00~Praha, Czech Republic 
}}

\title{Inertia Tames Fluctuations in Autonomous Stationary Heat Engines}

\author{Enrique P. Cital}\chau
\author{Viktor Holubec}\email{viktor.holubec@mff.cuni.cz}\chau
\date{\today}
\begin{abstract}
Thermodynamic uncertainty relations (TURs) provide fundamental constraints on the interplay between power fluctuations, entropy production, and efficiency in overdamped stationary autonomous heat engines. However, their validity in underdamped regimes remains limited and less explored. Here, we analytically and numerically study a physically realizable autonomous heat engine composed of two underdamped continuous degrees of freedom coupled to a two-level system. We show that this nonlinear setup can robustly violate TUR-based trade-offs by exploiting resonant coupling, effectively using one underdamped mode as an internal periodic drive. When this coupling is suppressed, the system recovers TUR-like bounds consistent with overdamped theory. Importantly, we demonstrate that the strongest suppression of current fluctuations occurs in a resonance regime that can be directly inferred from mean current measurements—a quantity typically much easier to access experimentally than fluctuations. Our results reveal new pathways to circumvent classical TUR constraints in underdamped systems and provide practical guidelines for designing efficient, precise microscopic engines and autonomous clocks.
\end{abstract}

\maketitle

\section{Introduction}

Since the days of Thomas Savery and Thomas Newcomen, who developed the first industrially relevant heat engines at the beginning of the eighteenth century, we have mastered the theory of converting abundant thermal energy into useful work on the macroscale~\cite{sandfort1979heat}. While important engineering challenges—such as decarbonization and pollution mitigation—remain to be solved~\cite{HEChallenges2023}, theoretical focus has shifted to heat engines operating at the microscale, where thermal fluctuations significantly affect their performance~\cite{Holubec_2022}. From the experimental perspective, this shift has been driven by advances in micromanipulation techniques, which have enabled the development of heat engines based on single Brownian~\cite{ColloidalReview} or quantum~\cite{Quantum_engines} particles. On the theoretical side, it is motivated by a deepening understanding of the fundamental principles of nonequilibrium statistical physics~\cite{Seifert_2012,SeifertReview2019,Holubec_2022}.

This branch of statistical physics is now known as stochastic thermodynamics~\cite{sekimoto2010stochastic, Seifert_2012}. In addition to the fluctuation theorems~\cite{Seifert_2012}, which generalize the second law of thermodynamics to small systems, one of its key results is the thermodynamic uncertainty relations (TURs)~\cite{horowitz2020thermodynamic}, which constrain the precision of thermodynamic currents by the amount of entropy produced. In its original form, the TUR reads~\cite{barato2015thermodynamic,Pietzonka2017}
\begin{equation}
\frac{\mathrm{Var}[j(t)]}{\langle j(t) \rangle^2} \geq \frac{2 k_B}{\sigma t},\label{eq:TUR}
\end{equation}
where
\begin{equation}
j(t) = \lim_{t_0 \to \infty} \frac{1}{t} \int_{t_0}^{t_0 + t} dt'\, \omega(t')
\label{eq:current_def}
\end{equation}
denotes the stationary current, averaged over the measurement time \(t\), associated with an observable \(\alpha\), where \(\omega = \dot{\alpha}\) is its instantaneous velocity.
The quantity $\sigma t$ is the total entropy production over the time interval $t$, $k_B$ is Boltzmann’s constant, and all averages are taken over the stochastic noise. 

Since heat and work currents in microscopic heat engines are also thermodynamic currents, the TUR~\eqref{eq:TUR} can be applied to constrain their performance. Specifically, for engines operating in a time-independent steady state while in thermal contact with two reservoirs at temperatures \( T_+ \) and \( T_- \), with \( T_+ > T_- \), one obtains the so-called power-efficiency-constancy trade-off (PECT)~\cite{pietzonka2018universal}:
\begin{equation}
    \frac{\mathrm{Var}[P(t)]\, t}{ \langle P \rangle} \frac{\eta_{\rm{C}} - \eta}{T_- \eta} \ge 2,
    \label{eq:PEC}
\end{equation}
where \( \eta_{\rm C} = 1 - T_-/T_+ \) denotes the Carnot efficiency, \( \eta \) is the efficiency of the engine, \( \langle P \rangle \) its average power output, and \( \mathrm{Var}[P(t)] \) the variance of the power measured over a time interval of length \( t \), due to thermal fluctuations. The product \( \mathrm{Var}[P(t)]\, t \) is in Ref.~\cite{pietzonka2018universal} referred to as the constancy of the engine.
When valid, the PECT implies that efficient engines ($\eta \to \eta_{\rm C}$) must exhibit strongly fluctuating power output, with $\mathrm{Var}[P(t)] t /\langle P \rangle \to \infty$. It is therefore important to identify heat engine models for which the TUR and PECT do not apply.

The range of validity of TUR~\eqref{eq:TUR}, and consequently PECT~\eqref{eq:PEC}, is limited to dynamics that obey local detailed balance, in the sense that the entropy change in the bath associated with a transition between two system states is proportional to the logarithm of the ratio of the forward and backward transition rates between those states~\cite{SeifertReview2019}. While overdamped Brownian motion can be formulated within this framework~\cite{Dechant_2018, Dechant2018}, underdamped Brownian motion~\cite{Vu2019}, processes with time‑dependent driving~\cite{Koyuk_2019}, or quantum processes~\cite{PhysRevLett.126.010602} require a more elaborate definition of time‑reversed dynamics, leading to additional terms in the corresponding TURs that depend on information about the system dynamics beyond entropy production alone. This makes such generalized TURs less useful for practical applications, where measuring current fluctuations can otherwise provide an estimate of the minimal entropy production required to sustain the observed current via a TUR~\cite{horowitz2020thermodynamic}.

To overcome this practical limitation for underdamped systems, where the TUR can be easily shown to be invalid for short measurement times~\cite{Fischer2020}, it was suggested—based on strong numerical evidence—that the standard TUR~\eqref{eq:TUR} holds for underdamped systems at least in the long-time limit~\cite{Fischer2020}, where the upper bound on precision is set by free diffusion under a constant drift. However, recently, using a cleverly designed two-dimensional setup reminiscent of a periodically driven system, a counterexample of an underdamped system violating the long-time TUR was presented~\cite{Pietzonka2022}.

In this work, we investigate—both analytically and through Brownian dynamics simulations—a similar design and demonstrate how it can be used to construct autonomous, stationary heat engines that robustly violate both the thermodynamic uncertainty relation (TUR) and the power-efficiency-constancy trade-off (PECT). The proposed engines combine a two‑dimensional underdamped system with a two‑level system. As in the setup of Ref.~\cite{Pietzonka2022}, one of the underdamped degrees of freedom effectively functions as an internal periodic drive, suggesting a connection of the setup to cyclic heat engines~\cite{Holubec2018.121.120601}. The strongest violations of the TUR and PECT occur in the regime of resonant coupling between the underdamped degrees of freedom. In this regime, the TUR can be violated to a similar extent as in setups relying on quantum coherence~\cite{Meier2025}. Here, however, the effect arises from classical inertia, i.e., from the presence of terms proportional to the first time derivative of the velocity in the underdamped dynamical equations, making it more experimentally accessible and robust against environmental noise. When the degree of freedom acting as the internal drive is decoupled from the engine, the system’s performance saturates both the TUR and the PECT, in agreement with the predictions of Ref.~\cite{Fischer2020}. Our results may be applied to the development of precise autonomous clocks or reliable microscopic heat engines and offer a practical pathway for identifying optimal operating regimes based solely on mean current measurements.

The remainder of this paper is structured as follows: In Sec.~\ref{sec:model}, we introduce the model. Section~\ref{sec:TD} defines the thermodynamic quantities relevant to the proposed system. In Sec.~\ref{sec:TUR}, we analyze the TUR and PECT, presenting both analytical and numerical evidence for their violation. Section~\ref{sec:TURopt} explores the parameter regimes in which TUR violations are maximized, highlighting the role of resonant coupling. In Sec.~\ref{sec:experiment}, we propose a feasible experimental implementation based on a thermoresponsive microgel setup. We conclude in Sec.~\ref{sec:Conclusion}. Appendices~\ref{appx:intuitionU}, \ref{appx:j2}, and \ref{appx:xv} provide additional analytical insights into current stabilization, current variance, and the dynamics of the positional and velocity degrees of freedom, respectively.

\section{Model}
\label{sec:model}

{Our goal is to construct a model of a stationary heat engine that is not constrained by the long-time TUR. To this end, we draw inspiration from the model of classical pendulum clocks introduced in Ref.~\cite{Pietzonka2022}, which is known to violate the long-time TUR. We construct a slight modification of that model so that it can operate as an autonomous heat engine. Specifically,} we consider a system composed of two continuous degrees of freedom, $\alpha(t)$ and $x(t)$, and one discrete degree of freedom, $d(t)$. For simplicity, we assume that all variables are dimensionless. The variable $\alpha(t)$ lies on a unit circle and can be interpreted as an angular coordinate; $x(t)$ represents a standard positional degree of freedom, while $d(t) \in \{-, +\}$ is a dichotomous Markov process.

The continuous degrees of freedom evolve according to the following system of dimensionless, generally damped {(sometimes also referred to as underdamped)} Langevin equations (explicit function arguments are omitted for conciseness):
\begin{align}
\dot{\omega} &=  - \tau_{w}  - \frac{\partial U_d}{\partial \alpha} - \frac{\partial U}{\partial \alpha} - \gamma_{\alpha} \omega + \sqrt{2D_{\alpha}\gamma_{\alpha}^2}\,\xi_\alpha, \label{eq:om} \\
\dot{v} &= -\frac{\partial U}{\partial x} - \gamma_{x} v + \sqrt{2D_{x}\gamma_{x}^2}\,\xi_x,
\label{eq:v}
\end{align}
where $\omega = \dot{\alpha}$ and $v = \dot{x}$ {are the velocities corresponding to the two degrees of freedom. The ratios of the momentum changes $\dot{\omega}$ and $\dot{v}$ to the frictional forces $\gamma_x v$ and $\gamma_\alpha \omega$ quantify the importance of inertial effects in the system. We assume that inertia is non-negligible.} The terms $\xi_\alpha(t)$ and $\xi_x(t)$ are normalized, unbiased Gaussian white noises. The friction coefficients $\gamma_{\alpha}$ and $\gamma_{x}$ and the corresponding diffusivities $D_\alpha$ and $D_x$ are related via the fluctuation–dissipation relation {\( D_i = \frac{T_-}{\gamma_i} = \frac{1}{\gamma_i} \),
assuming that the thermal energy—and, in the chosen dimensionless units, also the bath temperature \(T_-\)—is equal to 1.} The force $\tau_w$ is constant. The components $U$ and $U_d$ of the overall  potential $V_d = U + U_d$ are given by
\begin{align}    
    U_d(\alpha) &= \tau_d \left|\alpha - \pi\right| + \delta_{d +}\Delta \epsilon,
    \label{eq:Ud}\\
    U(x,\alpha) &= \frac{1}{2}\kappa_{x} \,x^2+\frac{1}{2}\kappa_{\alpha x}\left[x-\lambda \sin\bigl(\Omega\,\alpha\bigr)\right]^2.
    \label{eq:UPitzonka}
\end{align}

The potential \( U_d \) depends on the discrete degree of freedom \( d \) through the coefficient \( \tau_d \) and the Kronecker delta \( \delta_{d+} \). The corresponding torque 
\begin{equation}
 -\partial_\alpha U_d = \tau_d\sgn\left(\sin \alpha\right)   
 \label{eq:Ftl}
\end{equation}
is determined by the coefficients $\tau_\pm$ and is independent of the second term in $U_d$.

\begin{figure}
    \centering
  \includegraphics[width=0.5\columnwidth]{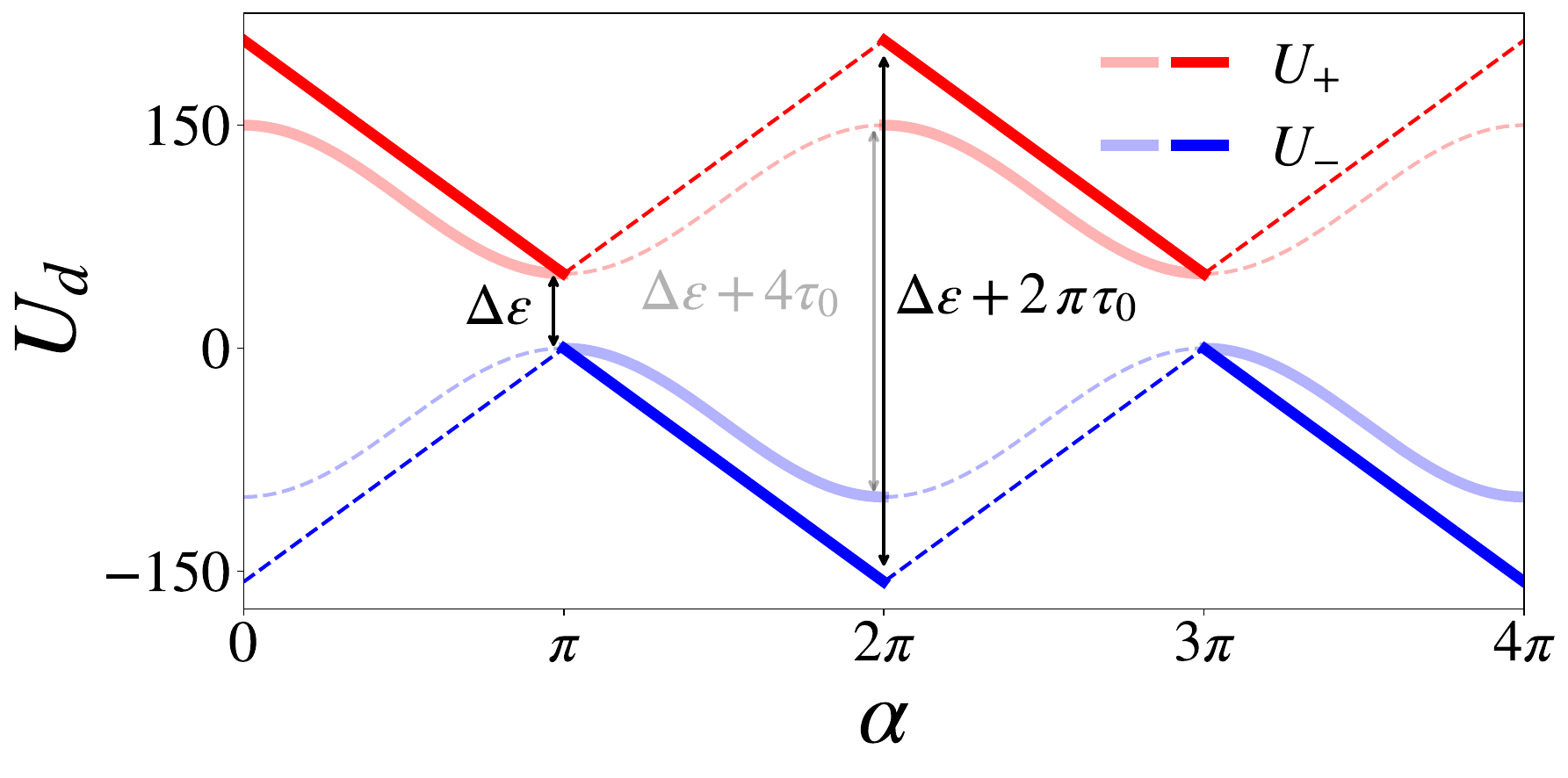}
    \caption{\emph{Operational principle of the heat engine.} Red and blue solid lines indicate regions of the angular coordinate $\alpha$ where the temperature~\eqref{eq:T} is hot ($T = T_+$) or cold ($T = T_-$), respectively. When the system parameters are tuned as described in the main text, the two-level system occupies the excited state $+$ with near-unit probability in the hot region and the ground state $-$ in the cold region (see Figs.~\ref{fig:1Dseries}c and~\ref{fig:1Dseries}f). As a result, the potential $U_d$ is, with high probability, given by the solid line, and the corresponding force—determined by the negative slope of this potential—is positive, generating a net current in $\alpha$-space (see Figs.~\ref{fig:1Dseries}a and~\ref{fig:1Dseries}b). Dashed lines indicate alternative potential configurations that are rarely realized. The sharp solid lines correspond to the piecewise linear potential~\eqref{eq:Ud} used in our theoretical analysis, while the transparent curves illustrate a possible experimental realization, proposed in Sec.~\ref{sec:experiment}, based on the sinusoidal potential~\eqref{eq:Udexp}. Our numerical results indicate that the TUR and PECT can be violated for the linear profile, but remain valid for the sinusoidal one.
}
    \label{fig:Ud}
\end{figure}

This second term, however, plays a role in the dynamics of $d$, which transitions between its two states according to $\alpha$-dependent transition rates:
\begin{align}
k_{-+}(\alpha) &= G\,e^{-\frac{1}{2 T(\alpha)}\left[U_+(\alpha)-U_-(\alpha)\right]}, \label{eq:k_+}\\
k_{+-}(\alpha) &= e^{-\frac{1}{2T(\alpha)}\left[U_-(\alpha)-U_+(\alpha)\right]}. \label{eq:k_-}
\end{align}
Interpreting the process $d$ as the state of a two-level system with a non-degenerate ground state and a $G$-fold degenerate excited state in thermal contact with a heat reservoir at temperature $T(\alpha)$,
these transition rates satisfy the detailed balance condition:
\(
\frac{k_{-+}}{k_{+-}} = G\,e^{-\frac{1}{T(\alpha)}(U_+ - U_-)}
\). In the following, we use a specific piecewise constant  temperature profile:
\begin{equation}
    T(\alpha) = 
    \begin{cases}
        T_+ & \text{for } \alpha \in [0, \pi), \\
        T_- & \text{for } \alpha \in [\pi, 2\pi),
    \end{cases}
    \label{eq:T}
\end{equation}
which switches between the cold temperature \(T_- = 1\), corresponding to the bath that governs the continuous degrees of freedom, and a hot temperature \(T_+ > 1\). The dynamics of the two‑level system drive the heat engine and generate a positive current~\eqref{eq:current_def} in the system. The engine’s operation is based on the ratchet potential $U_d$, as illustrated in Fig.~\ref{fig:Ud}.

The potential $U$ represents a linear combination of two harmonic traps with  stiffnesses $\kappa_{x}$ and $\kappa_{\alpha x}$, with minima at the origin and at $x = \lambda \sin\bigl(\Omega\,\alpha\bigr)$, respectively. This potential was shown in Ref.~\cite{Pietzonka2022} to suppress fluctuations of the current~\eqref{eq:current_def} associated with the degree of freedom \(\alpha(t)\) by reducing fluctuations in \(\omega(t)\), similarly to the model discussed in Appendix~\ref{appx:intuitionU}.

In Sec.~\ref{sec:experiment} we present a possible experimental implementation of this system. Now, we will briefly discuss its expected physics. 

The continuous degrees of freedom evolve isothermally under an equilibrium heat bath of unit thermal energy. In all the equations above, the potentials \(U_d\) (\(d = \pm\)) and \(U\)  may be substituted by a single potential \(V_d = U + U_d\). Hence, the system is driven out of equilibrium by two distinct mechanisms:

\emph{The non‐conservative torque} \(\tau_w\) in Eq.~\eqref{eq:om}, which induces a persistent counter-clockwise rotation of the angle \(\alpha\). For vanishing conservative force, $\partial_\alpha V_d = 0$, the stationary average current is $\langle \omega \rangle = - \tau_w/\gamma_\alpha$, as follows from Eq.~\eqref{eq:om} upon averaging and using the stationarity condition $\langle \dot{\omega} \rangle = 0$. 
  
\emph{Spatially varying temperature} \(T(\alpha)\), {which governs transitions in the two-level subsystem and synchronizes the continuous and discrete dynamics. When \(\sgn \tau_+ \neq \sgn \tau_-\), this synchronization can induce a nonzero stationary current}, driven by heat flux from the hot reservoir. The current is maximized when the two-level dynamics becomes deterministic—for example, when the system remains in state \(+\) for \(\alpha \in [0, \pi)\) and in state \(–\) for \(\alpha \in [\pi, 2\pi)\) if \(\tau_+ > 0\) and \(\tau_- < 0\). In this case, the torque in Eq.~\eqref{eq:Ftl} remains positive for all \(\alpha\). The opposite configuration—where \(\tau_+ < 0\), \(\tau_- > 0\), \(d = -\) for \(\alpha \in [0, \pi)\), and \(d = +\) for \(\alpha \in [\pi, 2\pi)\)—produces a negative force of the same magnitude, propelling the system in the reversed direction.
  
We will operate the system as a heat engine, using thermal energy from the hot bath at temperature \( T_+ \) to induce a stationary current \( \langle \omega \rangle \) against the torque \( \tau_w \), thus producing power \( P = \tau_w \langle \omega \rangle \). Therefore, we choose \( \tau_+ > 0 \) and \( \tau_- < 0 \). To minimize current fluctuations and simplify analytical calculations, we moreover set
\begin{equation}
    \tau_+ = - \tau_- = \tau_0 > 0 \label{eq:tau0}
\end{equation}
{which yields a constant torque
\begin{equation}
 -\partial_\alpha U_d = \tau_0.   
 \label{eq:FtlOUrRegime}
\end{equation}
}

\begin{figure}
    \centering
  \includegraphics[width=0.76\textwidth]{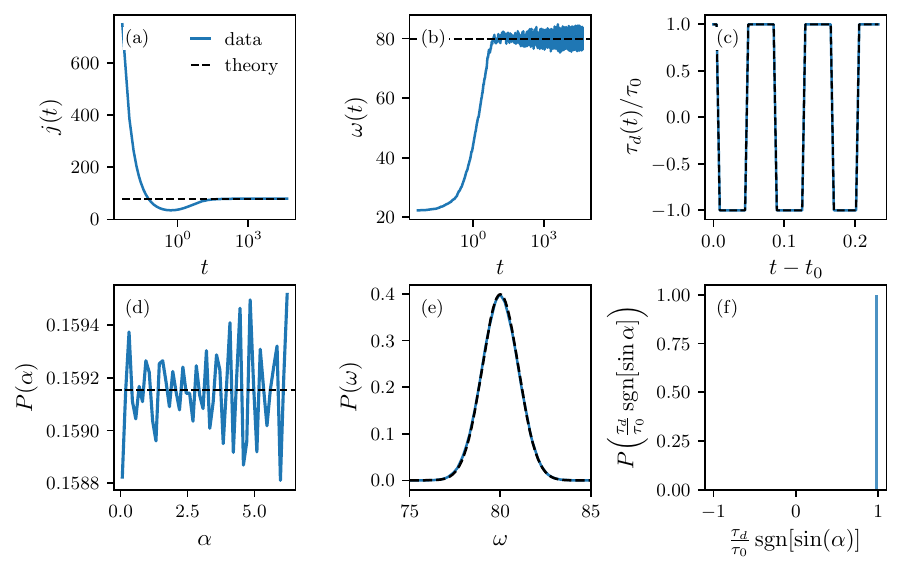}
    \caption{\emph{Dynamics of the system for \(\partial_\alpha U = 0\).}  
(a) Current \(j(t)\) in Eq.~\eqref{eq:current_def},  
(b) the corresponding angular velocity \(\omega(t)\), defined in Eq.~\eqref{eq:om},  
(c) the state of the two-level system, expressed as \(\tau_d(t)/\tau_0\), all shown as functions of time.  
(d) Probability density of the angular coordinate \(\alpha(t)\), corresponding to \(\omega(t)\) in (b).  
(e) Probability density of \(\omega(t)\).  
(f) Probability density of \(\tau_d \, \text{sgn}[\sin(\alpha)]/\tau_0\), illustrating the deterministic dynamics of the two-level system in the given parameter regime.  
Solid blue lines represent results from Brownian dynamics simulations, while black dashed lines show theoretical predictions:  
(a–b) \( j(t)\, t = \omega_t\, t \) and \( \omega(t) = \omega_t \), with \( \omega_t \) given in Eq.~\eqref{eq:omT};  
(c) \( \tau_d(t)/\tau_0 = \text{sgn}[\sin(\alpha(t))] \);  
(d) \( P(\alpha) = 1/(2\pi) \);  
(e) \( P(\omega) \) is given by Eq.~\eqref{eq:Pomega}, with the mean shifted to \( \omega_t \).
In panel (d), \(t_0 = t_{max} - 3 \times 2\pi / \omega_t\).  
All parameters, with the exception of \(\kappa_{\alpha x} = 0\), are listed in Tab.~\ref{tab:parameters}.
}
    \label{fig:1Dseries}
\end{figure}

The two-level system will reside in state \( + \) for \( \alpha \in [0, \pi) \) and in state \( - \) for \( \alpha \in [\pi, 2\pi) \) when \( k_+(\alpha) \gg k_-(\alpha) \) for \( \alpha \in [0, \pi) \), and \( k_+(\alpha) \ll k_-(\alpha) \) for \( \alpha \in [\pi, 2\pi) \). Under the condition~\eqref{eq:tau0}, the ratio of transition rate in Eqs.~\eqref{eq:k_+} and \eqref{eq:k_-} is given by
\[
\frac{k_{-+}}{k_{+-}}
= G \exp\left(- \frac{\Delta \epsilon + 2\tau_0 |\alpha - \pi|}{ T(\alpha)}\right).
\]
The ratio decreases with the energy difference \( U_+(\alpha) - U_-(\alpha) \), depicted in Fig.~\ref{fig:Ud}, and increases with temperature and \( G \).  
The limiting regime described above imposes the following conditions on the model parameters:
\begin{equation}
\begin{aligned}
G \exp\left(- \frac{\Delta \epsilon + 2\tau_0 |\alpha - \pi|}{ T_+}\right) &\gg 1, && \alpha \in [0, \pi), \\
G \exp\left(- \frac{\Delta \epsilon + 2\tau_0 |\alpha - \pi|}{ T_-}\right) &\ll 1, && \alpha \in [\pi, 2\pi).
\end{aligned}
\label{eq:par_reg}
\end{equation}
 The first condition requires large enough \( G \) and \( T_+ \), and the second one large enough \( \Delta \epsilon \). For specific required ratios of the rates, these parameters can be calculated explicitly.
In our numerical analysis below, we use the parameters in Tab.~\ref{tab:parameters}, which yield $\min_{\alpha \in [0,\pi)} k_{-+}/k_{+-} \approx 9 \times 10^9$ and $\max_{\alpha \in [\pi,2\pi)} k_{-+}/k_{+-} = 2 \times 10^{-12}$, effectively fulfilling the requirement on deterministic time-evolution of the two-level system, as verified in Figs.~\ref{fig:1Dseries}c and \ref{fig:1Dseries}f. 

When \(T(\alpha) = T_-\) and \(\tau_w = 0\), the system relaxes to thermal equilibrium. The stationary joint probability density \(P_d(\alpha,\omega,x,v)\) for \(\alpha\), \(\omega\), \(x\), \(v\), and \(d\) is thus given by the Boltzmann distribution
\begin{equation}
    P_d(\alpha,\omega,x,v) = \frac{1}{Z} \exp\left[-\frac{V_d(\alpha, x)}{T_-}-\frac{\omega^2+v^2}{2T_-}\right],
    \label{eq:PallEq}
\end{equation}
where \(Z\) is the partition function. Equilibrium probability densities for the individual variables can be obtained by marginalizing the full distribution \( P_d(\alpha, \omega, x, v) \) (for conciseness, we denote all probability densities for continuous degrees of freedom by \( P \), distinguishing them only by their variables). In particular, \( P(\omega) \) is, in equilibrium, given by
\begin{equation}
    P(\omega) = \frac{1}{\sqrt{2\pi T_-}} \exp\left(-\frac{\omega^2}{2T_-}\right).
    \label{eq:Pomega}
\end{equation}

\begin{table}[h!]
\centering
\begin{tabular}{|c|c|c|c|c|c|c|c|c|c|c|c|c|c|}
\hline
$G$ & $T_-$ & $T_+$ & $\Delta \epsilon$ & $\tau_0$& $\tau_w$ & $\gamma_{\alpha}$ & $\gamma_x$ & $D_{\alpha}$ & $D_x$ & $\kappa_x$ & $\kappa_{\alpha x}$ & $\Omega$ & $\lambda$ \\
\hline
$10^{10}$ & $1$ & $10^3$ & $50$ & $50$ & $10$& $0.5$ & $0.5$ & $2$ & $2$ & $500$ & $100$ & $1$ & $1$ \\
\hline
\end{tabular}
\caption{\emph{List of model parameters used.} Brownian dynamics simulations were performed using the Verlet-type algorithm described in Ref.~\cite{Grønbech-Jensen01042013}, with a timestep of \(dt = 5 \times 10^{-7}\). Under these conditions, a single simulation run with duration \(t_{\text{max}} = 45000\) (Figs.~\ref{fig:1Dseries}--\ref{fig:2DTUR}) or \(t_{\text{max}} = 9000\) required approximately 8 hours and 1.5 hours of computation time, respectively, on a desktop PC equipped with an AMD Ryzen 7 5700X 8-core processor (16 threads), featuring a base clock speed of 3.4~GHz and a boost frequency of up to 4.7~GHz.
}
\label{tab:parameters}
\end{table}

\section{Thermodynamics}
\label{sec:TD}

Let us now describe the thermodynamics of the engine {in the parameter regime described in the preceding section. For \( \partial_\alpha U = 0 \), where the dynamics of \(\alpha\) are decoupled from those of \(x\), the total torque in Eq.~\eqref{eq:om} is constant and is given by \( \tau_0 - \tau_w \).} Consequently, {the stationary average $\langle \omega \rangle$ of current  in Eq.~\eqref{eq:current_def}} is given by
\begin{equation}
    \langle j \rangle = \langle \omega \rangle = \omega_{t} 
    = \frac{\tau_0 - \tau_w}{\gamma_\alpha}
    \qquad \text{for } \partial_\alpha U = 0 \, .
    \label{eq:omT}
\end{equation}
({Here and below, $\langle \bullet \rangle$ denotes the average over the noise taken in the steady state. 
In our Brownian dynamics simulations, we proceed as follows. First, we generate a single long trajectory $\omega(t_{\rm max})$ and discard the initial relaxation period. We then divide the remaining trajectory of length $t_{\rm m}$ into $t_{\rm m}/t$ segments and calculate the time-averaged current defined in Eq.~\eqref{eq:current_def} for each segment. Finally, we average over these samples of the time-averaged current, obtaining $\langle j \rangle$. Higher moments of $j$ are computed analogously.}) 

Under isothermal conditions, the current induced by a constant force in a periodic potential is always smaller than the current induced by the same force in the absence of any potential~\cite{REIMANN200257}. Hence, for a nonzero \( \partial_\alpha U \), $\omega_{t}$ serves as an upper estimate for the average current:
\begin{equation}
\langle j \rangle = \langle \omega \rangle \le \omega_{t} \qquad \text{for } \partial_\alpha U \neq 0 \, .
\label{eq:j}
\end{equation}
The maximum can be saturated by appropriately fine-tuning the parameters of the potential $U$, as shown in Fig.~\ref{fig:optimisation} in Sec.~\ref{sec:TURopt}.

{The dynamics of the degree of freedom \( x \) is irreversible due to the coupling to $\alpha$. The corresponding heat flux into the cold bath is \( q_x = \left\langle \left(\gamma_x v - \sqrt{2D_x\gamma_x}\xi_x\right)v \right\rangle = - \langle \partial_x U v \rangle \)~\cite{sekimoto2010stochastic}, where we used Eq.~\eqref{eq:v} and the stationarity condition $\langle \dot{v} v\rangle = \dot{\langle v^2 \rangle}/2 = 0$. Similarly, the heat flux into the cold bath related to the current in \( \alpha \) is given by \(q_\alpha = \left\langle \left(\gamma_\alpha \omega - \sqrt{2D_\alpha \gamma_\alpha^2}\xi_\alpha(t) \right) \omega \right\rangle = - \left\langle \partial_\alpha U \omega \right\rangle + (\tau_0 - \tau_w) \left\langle \omega \right\rangle \), where we used Eq.~\eqref{eq:FtlOUrRegime}. The overall heat flux into the cold bath from continuous degrees of freedom is thus $q_x + q_\alpha = (\tau_0 - \tau_w) \langle \omega \rangle$, because, from stationarity, $\langle \dot{U} \rangle =  \langle \partial_x U v + \partial_\alpha U \omega \rangle = 0$. Since the continuous degrees of freedom couple only to the cold thermal reservoir at temperature $T_-$, the corresponding entropy production reads
\begin{equation}
 \sigma_c  =  \frac{(\tau_0 - \tau_w) \langle \omega \rangle}{T_-} 
 \le \frac{\gamma_{\alpha} \omega_{t}^2}{T_-},
 \label{eq:sigmac}
\end{equation}
where the upper estimate holds for $\partial_\alpha U = 0$ and we used Eqs.~\eqref{eq:omT} and \eqref{eq:j}.}

From Fig.~\ref{fig:Ud}, one can deduce that, in the chosen `deterministic' parameter regime, the two-level system absorbs per one revolution of \( \alpha \) the energy \( U_+(0) - U_-(0) = \Delta \epsilon + 2 \pi \tau_0 \) from the hot reservoir and releases the energy \( \Delta \epsilon \) to the cold reservoir. One revolution in the steady state takes on average the time $2\pi/ \langle \omega \rangle$. Hence, the
heat influx $q_h = (\Delta \epsilon + 2\pi \tau_0)\langle \omega \rangle/2\pi$ and the operation of the two-level system is accompanied by the entropy production 
\begin{equation}
    \sigma_{2L} = \left( \frac{\Delta \epsilon}{T_-} - \frac{\Delta \epsilon + 2\pi \tau_0}{T_+} \right) \frac{\langle \omega \rangle}{2\pi} \le 
    \left( \frac{\Delta \epsilon}{T_-} - \frac{\Delta \epsilon + 2\pi \tau_0}{T_+} \right) \frac{\omega_{t}}{2\pi}.
    \label{eq:sigma2L}
\end{equation}
Total entropy production in the system is $\sigma = \sigma_c + \sigma_{2L}$.

The mean output power, $\langle P \rangle$, and efficiency, $\eta$, of the engine then read
\begin{align}
    \langle P\rangle &= \tau_w \langle \omega \rangle \le \tau_w \omega_{t},\\
    \eta & = \frac{\langle P\rangle}{q_h} = 
    \frac{2\pi \tau_w }{\Delta \epsilon + 2\pi \tau_0}.
    \label{eq:eta}
\end{align}
{Note that, although we provide only an upper bound for the power—which saturates in the regime where \(\alpha\) is independent of \(x\)—the prediction for the efficiency is exact in the parameter regime described in the preceding section, because the generally unknown average current $\langle \omega \rangle$ cancels between the numerator and the denominator}. The second law bound on efficiency, $\eta \le 1 - T_-/T_+$ gives the condition $T_-/T_+ \le [\Delta \epsilon + 2\pi (\tau_0-\tau_w)]/(\Delta \epsilon + 2\pi \tau_0)$, under which the system operates as a heat engine for positive values of parameters in Eq.~\eqref{eq:eta}.

\begin{figure}
    \centering
  \includegraphics[width=0.5\columnwidth]{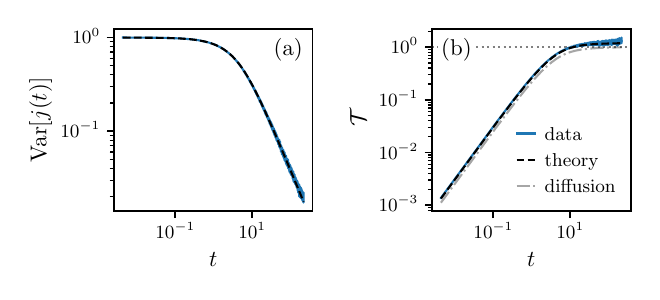}
    \caption{\emph{Time-resolved TUR for \(\partial_\alpha U = 0\):}  
(a) Current variance and (b) \(\mathcal{T}\) from Eq.~\eqref{eq:T} as functions of the measurement duration \(t\).  
Blue solid lines correspond to Brownian dynamics simulations, while black dashed lines show theoretical predictions from Eqs.~\eqref{eq:varj} (upper bound) and \eqref{eq:TFree}.  
The grey dot-dashed line in (b) depicts \(\mathcal{T}\) for free diffusion under constant drift, given by Eq.~\eqref{eq:TFree} with \(\sigma_{2L} = 0\).  
The TUR in Eq.~\eqref{eq:T} is violated when \(\mathcal{T}\) in panel (b) falls below the grey dotted line marking the threshold \(\mathcal{T} = 1\).
{The breakdown of the TUR at short times is expected, as it is known to be invalid in this regime~\cite{Fischer2020}.}}
    \label{fig:1DTUR}
\end{figure}

\begin{figure}
    \centering
  \includegraphics[width=1.0\textwidth]{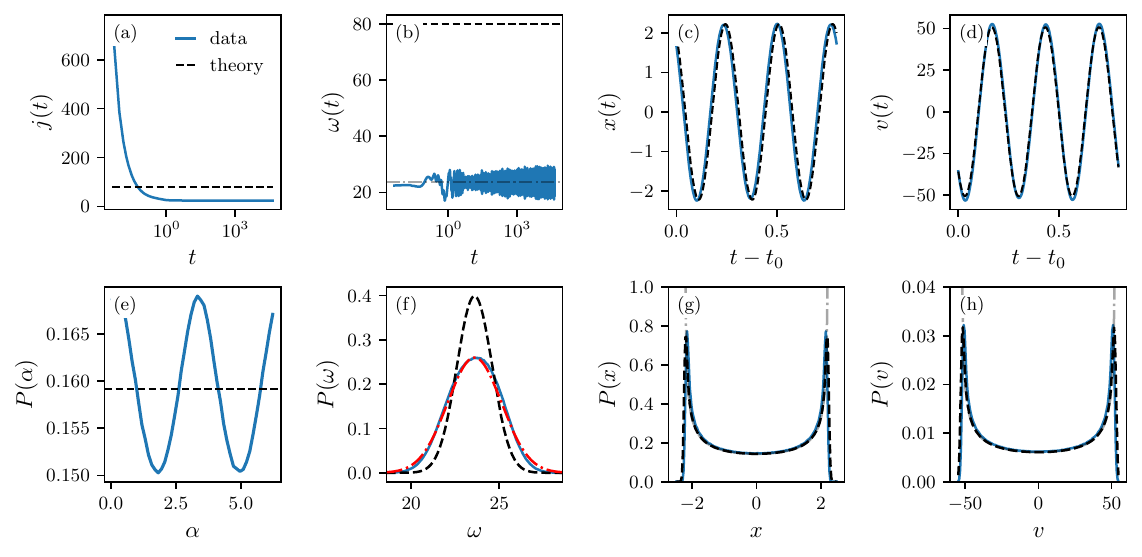}
    \caption{\emph{Dynamics of the full system.}  
(a) Current \( j(t) \) in Eq.~\eqref{eq:current_def};  
(b) corresponding angular velocity \( \omega(t) \), defined in Eq.~\eqref{eq:om};  
(c) position \( x(t) \); and  
(d) velocity \( v(t) \), defined in Eq.~\eqref{eq:v},  
all shown as functions of time.  
(e) Probability density of the angular coordinate \( \alpha(t) \), corresponding to \( \omega(t) \) in (b).  
(f–h) Probability densities of \( \omega(t) \), \( x(t) \), and \( v(t) \), corresponding to panels (b), (c), and (d), respectively.  
Solid blue lines represent results from Brownian dynamics simulations, while black dashed lines denote theoretical predictions:  
(a–b) \( j(t)\, t = \omega_t\, t \) and \( \omega(t) = \omega_t \), with \( \omega_t \) given in Eq.~\eqref{eq:omT};  
(c–d) \( x(t) \) and \( v(t) \) given by Eqs.~\eqref{eq:xt} and \eqref{eq:vt};  
(e) \( P(\alpha) = 1/(2\pi) \);  
(f) \( P(\omega) \) is given by Eq.~\eqref{eq:Pomega}, with the mean shifted to \( \langle \omega \rangle \) [dot–dashed grey line in (b)]; additionally, the red dot–dashed line shows the same distribution with the variance rescaled as \( T_- \to 2.35 \);
(g–h) \( P(x) \) and \( P(v) \) are given by Eqs.~\eqref{eq:px}, \eqref{eq:pv}, and \eqref{eq:PConv}. Additionally, the grey dot-dashed lines show the distributions from Eqs.~\eqref{eq:px} and \eqref{eq:pv} without the convolution in Eq.~\eqref{eq:PConv} with the noise kernel.
In panels (c–d), \( t_0 = t_{\mathrm{max}} - 3 \times 2\pi / \omega_t \).  
All parameters are listed in Tab.~\ref{tab:parameters}.
 }
    \label{fig:2Dseries}
\end{figure}

\begin{figure}
    \centering
  \includegraphics[width=0.5\columnwidth]{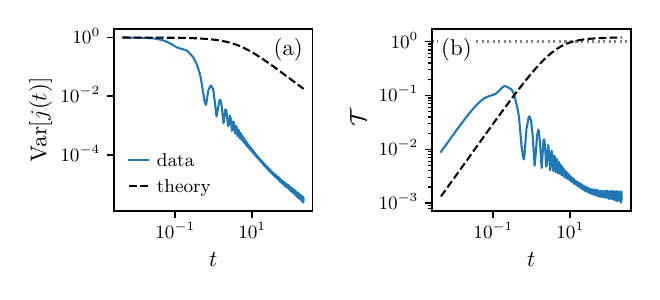}
    \caption{\emph{Time-resolved TUR in the full system ($\partial_\alpha U \neq 0$):}  
(a) Current variance and (b) \(\mathcal{T}\) from Eq.~\eqref{eq:T} as functions of the measurement duration \(t\).  
Blue solid lines are obtained from Brownian dynamics simulations, while black dashed lines represent theoretical predictions from Eqs.~\eqref{eq:varj} (upper bound) and \eqref{eq:TFree}, which apply to \(\partial_\alpha U\) when the \(\alpha\)-dynamics is independent of \(x\).  
The TUR in Eq.~\eqref{eq:T} is violated when \(\mathcal{T}\) in panel (b) drops below the grey dotted line, which marks the threshold \(\mathcal{T} = 1\). {In this case, the expected breakdown of the TUR at short times~\cite{Fischer2020} extends to arbitrary times, similarly to the classical pendulum clocks considered in Ref.~\cite{Pietzonka2022}.}
}
    \label{fig:2DTUR}
\end{figure}

\begin{figure*}
    \centering
  \includegraphics[width=1.0\textwidth]{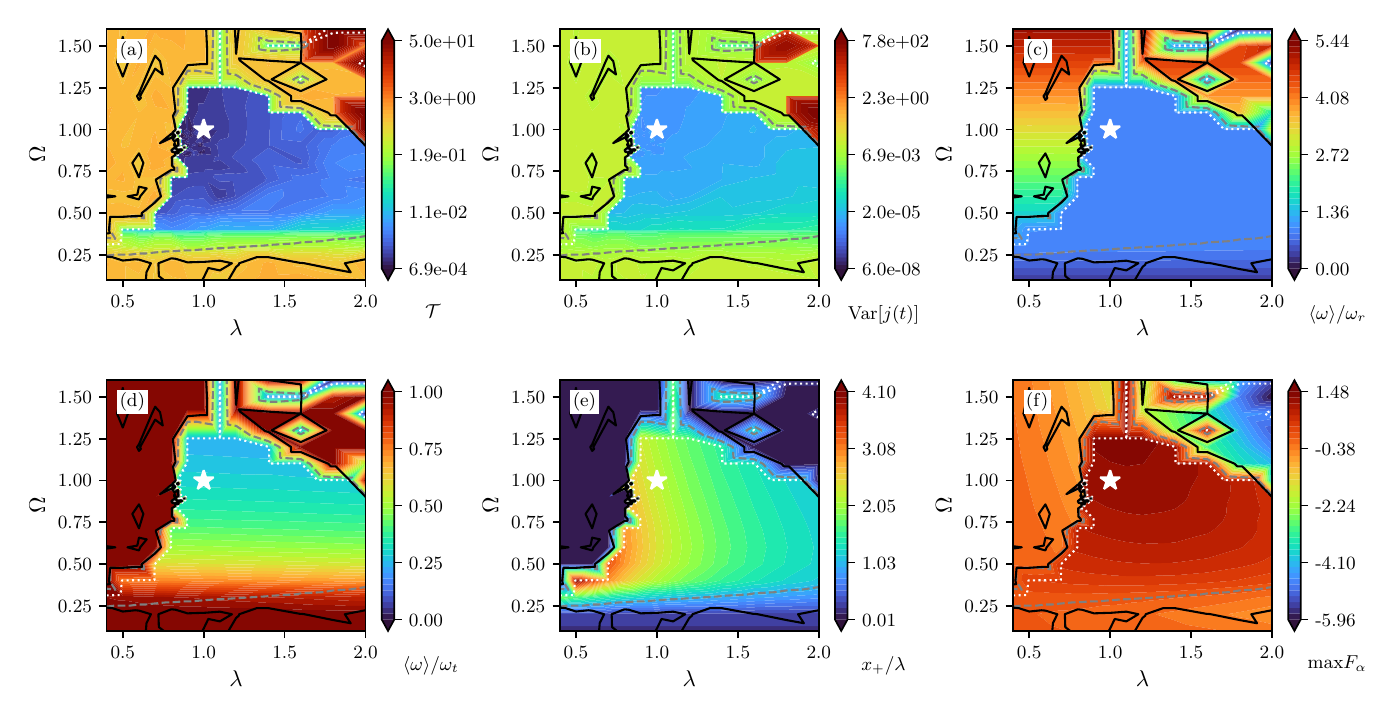}
    \caption{\emph{Phase diagram for the TUR:}  
(a) Asymptotic value \(\mathcal{T}_\infty\) of \(\mathcal{T}\) from Eq.~\eqref{eq:T}, evaluated at time \(t = 10^2\), as a function of \(\Omega\) and \(\lambda\), with all other parameters as specified in Tab.~\ref{tab:parameters}. 
Panels (b)--(f) display corresponding quantities computed for the same parameters:  
(b) the current variance,  
(c) the average current (angular frequency) $\langle \omega \rangle$ relative to the resonance frequency $\omega_r$ from Eq.~\eqref{eq:omR},  
(d) $\langle \omega \rangle$ relative to the average current $\omega_t$ for $\partial_\alpha U = 0$ from Eq.~\eqref{eq:omT},  
(e) the amplitude $x_+$ of oscillations in $x$, as predicted by Eq.~\eqref{eq:xp}, normalized by $\lambda$, and  
(f) an estimate $\max F_\alpha = \lambda \Omega (x_+ - \lambda)$ of the magnitude of the force $-\partial U_\alpha$ taming the current fluctuations. Entropy production $\sigma = \sigma_c + \sigma_{2L}$ is proportional to average current $\langle \omega \rangle$ [see Eqs.~\eqref{eq:sigmac} and \eqref{eq:sigma2L}] and thus its relative magnitude can be read from panel (d). The white star in all panels indicates the parameter regime used in Figs.~\ref{fig:2Dseries} and \ref{fig:2DTUR}, corresponding to \(\Omega = \lambda = 1\). In all panels, the dashed black line marks \(\mathcal{T}_\infty = 1\), the dot-dashed grey line indicates \(x_+ / \lambda = 1\) and \(\max F_\alpha = 0\), and the dotted white line corresponds to \(\langle \omega \rangle / \omega_r = 1\).
    }
    \label{fig:optimisation}
\end{figure*}

\section{Thermodynamic uncertainty relations}
\label{sec:TUR}

To investigate the validity of the TUR~\eqref{eq:TUR}, we rewrite it in the form
\begin{equation}
\mathcal{T} \equiv \frac{\mathrm{Var}[j(t)] \, \sigma t}{2 \langle j(t) \rangle^2}
\ge 1.
\label{eq:T}
\end{equation}
In the present case, the fluctuating output power \(P(t)\) of the engine is proportional to the current \(j(t)\): \(P(t) = \tau_w j(t)\). In such a case, \(\mathcal{T}\) in Eq.~\eqref{eq:T} can be directly rewritten in the form of the PECT~\eqref{eq:PEC}, in terms of the efficiency \(\eta\), mean power \(\langle P(t) \rangle = \tau_w \langle j(t) \rangle\), and power variance \(\mathrm{Var}[P(t)] = \tau_w^2 \mathrm{Var}[j(t)]\)~\cite{pietzonka2018universal}. 
Specifically, using \(\eta_{\rm C} = 1 - T_-/T_+\), \(\eta =\langle P \rangle/q_h\)
, and \(\sigma = q_c/T_- - q_h/T_+\), where \(q_c\) and \(q_h\) are the mean heat fluxes into the cold bath and out of the hot bath, respectively, one obtains
\begin{equation}
    \mathcal{T} = \frac{\mathrm{Var}[P(t)]\, t}{\langle P \rangle} \frac{\eta_{\rm{C}}-\eta}{2 T_- \eta}.
    \label{eq:T2}
\end{equation}
Thus, in our case, violation of the ‘classical’ TUR in Eq.~\eqref{eq:TUR} is equivalent to the violation of the trade-off PECT in Eq.~\eqref{eq:PEC}.

For \(\partial_\alpha U = 0\), the process \(\alpha(t)\) is a free diffusion with a constant drift \(\tau_0 - \tau_w\). The stationary variance of the corresponding current~\eqref{eq:current_def} is then given by Eq.~\eqref{eq:varjappx} in Appendix~\ref{appx:j2}~\cite{Fischer2020}, as verified in Fig.~\ref{fig:1DTUR}a. The potential \(U\) is crafted to tame the current fluctuations~\cite{Pietzonka2022}. Hence, when \(\partial_\alpha U \neq 0\), we expect that the current fluctuations are suppressed, leading to a smaller variance,
\begin{equation}
    \mathrm{Var}[j(t)] \le \frac{2 D_{\alpha}}{t^2} \left( t + \frac{e^{-\gamma_{\alpha} t} - 1}{\gamma_{\alpha}} \right).
\label{eq:varj}
\end{equation}
For the parameters in Tab.~\ref{tab:parameters}, this conjecture is verified in Fig.~\ref{fig:2DTUR}a.

Altogether, we obtain the following predictions for the TUR. For \(\partial_\alpha U = 0\), the process \(\alpha(t)\) is independent of \(x(t)\) and, in the considered parameter regime, corresponds to a free diffusion with constant drift. Equations~\eqref{eq:j}, \eqref{eq:sigmac}, \eqref{eq:sigma2L}, and \eqref{eq:varj} yield
\begin{equation}
\mathcal{T} \equiv \frac{\mathrm{Var}[j(t)] \, \sigma}{2 \langle j(t) \rangle^2}
= \left[ 1 + \frac{e^{-\gamma_{\alpha} t} - 1}{\gamma_{\alpha} t} \right] \left( 1 + \frac{D_\alpha \sigma_{2L}}{\langle \omega \rangle^2} \right).
\label{eq:TFree}
\end{equation}
At short times, \(e^{-\gamma_{\alpha} t} \approx 1 - \gamma_\alpha t\), and thus \(\mathcal{T} \approx 0\). In this regime, the TUR~\eqref{eq:T} does not hold, as discussed in Ref.~\cite{Fischer2020}.
At long times, \(\mathcal{T} = 1 + \frac{D_\alpha \sigma_{2L}}{\langle \omega \rangle^2}\), and the TUR saturates (\(\mathcal{T} = 1\)) in the parameter regime where the entropy production in the continuous degree of freedom \(\alpha(t)\), given by Eq.~\eqref{eq:sigmac}, dominates over that in the two-level system, given by Eq.~\eqref{eq:sigma2L}, i.e., when \(\sigma_c \gg \sigma_{2L}\) and the entire system can be effectively described as free diffusion under a constant drift~\cite{Fischer2020}.  
For the parameters used in our numerical demonstration (Tab.~\ref{tab:parameters}), we find \(\sigma_c / \sigma_{2L} \approx 5\), resulting in a long-time value of \(\mathcal{T} \approx 1.2\).  
These results are confirmed numerically in Fig.~\ref{fig:1DTUR}. Figure~\ref{fig:1Dseries} additionally shows that the relaxation time of the angular velocity \( \omega(t) \) is on the order of 10 (Fig.~\ref{fig:1Dseries}b), and in the steady state, \( P(\omega) \) is given by the equilibrium distribution~\eqref{eq:Pomega} with the mean shifted to \( \omega_t \) (Fig.~\ref{fig:1Dseries}c). On the other hand, the probability distribution for \( \alpha \) in Fig.~\ref{fig:1Dseries}d is (up to fluctuations caused by the finite runtime \( t_{\mathrm{max}} \) of our simulations) flat, and thus very far from the equilibrium distribution predicted by Eq.~\eqref{eq:PallEq}.

For \(\partial_\alpha U \neq 0\), we must use the inequalities in Eqs.~\eqref{eq:j}, \eqref{eq:sigmac}, \eqref{eq:sigma2L}, and \eqref{eq:varj}. However, since we only have upper bounds for the individual quantities entering \(\mathcal{T}\), we cannot provide a definitive analytical estimate and must instead rely on numerical simulations. Results of our simulations, performed in the parameter regime listed in Tab.~\ref{tab:parameters}, are shown in Figs.~\ref{fig:2Dseries} and \ref{fig:2DTUR}. They clearly demonstrate that the potential \( U \) suppresses current fluctuations even in the autonomous heat engine case: even though the current shown in Fig.~\ref{fig:2Dseries}b is about four times smaller than for \( \partial_\alpha U = 0 \), its variance in Fig.~\ref{fig:2DTUR}a is smaller by several orders of magnitude. As a result, \( \mathcal{T} \) in Fig.~\ref{fig:2DTUR}b remains significantly below 1 at all measurement times \( t \), and the TUR~\eqref{eq:T} is violated for all times. Compared to the case with \( \partial_\alpha U = 0 \), the distribution of \( \omega \) in Fig.~\ref{fig:2Dseries}f is no longer given by the shifted equilibrium distribution~\eqref{eq:Pomega} (dashed black line), nor is it possible to fit it perfectly with a Gaussian (red dot-dashed line). Furthermore, the distribution for \( \alpha \) in Fig.~\ref{fig:2Dseries}e is still approximately constant; nevertheless, it exhibits deviations from this shape that are larger than the fluctuations caused by the finite runtime of our simulations.

\section{Parameter Regimes for TUR Violation}
\label{sec:TURopt}

In Appendix~\ref{appx:intuitionU}, we provide a simple model that offers intuition for the mechanism by which the potential \( U \) reduces fluctuations. When \( \omega(t) \) is at its average value, the motion in \( x \) and \( \alpha \) is optimally coupled, and the system’s trajectory closely follows the minimum of the potential, as illustrated in Fig.~\ref{fig:MODEL_SCHEME}c. When \( \omega(t) \) deviates from its average—either increasing or decreasing—this optimal coupling is disrupted, and the potential \( U \) exerts a restoring force on \( \omega \), driving it back toward the optimal state and thereby stabilizing the current. Ref.~\cite{Pietzonka2022} showed that this mechanism suppresses fluctuations strongly enough that the TUR is violated only when stochastic trajectories penetrate deeply into the potential well around its minimum. Let us now investigate the conditions under which the present system operates in this parameter regime.

Results from our Brownian dynamics simulations, shown in Fig.~\ref{fig:2Dseries}b, suggest that it is reasonable to assume the angular velocity is approximately constant. Thus, \(\alpha(t)\) in Eq.~\eqref{eq:v} can be approximated as \(\alpha(t) = \langle \omega \rangle t\). Under this approximation, and neglecting noise, Eq.~\eqref{eq:v} describes a harmonic oscillator driven by a sinusoidal force, which can be solved analytically, as shown in Appendix~\ref{appx:xv}. The resulting analytical expressions agree well with the Brownian dynamics simulation data, both for the time series of the position \(x(t)\) and velocity \(v(t)\), as well as for the corresponding probability densities, as shown in Figs.~\ref{fig:2Dseries}c, d, g, and h.

When valid, the analysis in Appendix~\ref{appx:xv} shows that \( x(t) \) oscillates with frequency \( \Omega \langle \omega \rangle \) and amplitude \( x_+ \), as given in Eq.~\eqref{eq:xp}.  
Neglecting fluctuations, the restoring force exerted by the potential \( U \) on the angular velocity \( \omega(t) \) vanishes when \( \lambda = x_+ \).  
Whenever \( x_+ \) deviates from this value, the potential \( U \) exerts a systematic restoring force on \( \alpha(t) \), guiding its dynamics toward the ideal trajectory \( \alpha(t) = \langle \omega \rangle t \).  
This restoring effect is maximal when \( x_+ \) attains its largest possible value, which occurs when \( \langle \omega \rangle \) matches the resonance frequency \( \omega_r \) defined in Eq.~\eqref{eq:omR}.  
Based on this reasoning, one may conjecture that the thermodynamic uncertainty relation (TUR) is most strongly violated when \( \langle \omega \rangle = \omega_r \).

We validate this conjecture in Fig.~\ref{fig:optimisation}a by plotting the asymptotic value $ \mathcal{T}_\infty = \lim_{t \to \infty} \mathcal{T} $, which we approximate in practice as $ \mathcal{T}_\infty \approx \mathcal{T}(10^2) $. The corresponding current variance is shown in Fig.~\ref{fig:optimisation}b. Both quantities are plotted as functions of $\Omega$ and $\lambda$, and exhibit a sharp transition as these parameters are varied: from large values, where the TUR holds or is only weakly violated ($ \mathcal{T} \sim 1 $), to small values, where the TUR is strongly violated ($ \mathcal{T} \sim 10^{-3} $).
Figure~\ref{fig:optimisation}c shows that the upper boundary (large $\Omega$) of the regime in which the TUR is strongly violated is precisely determined by the resonance condition $ \langle \omega \rangle = \omega_r $ (white dotted line). The lower boundary (small $\Omega$) is approximately set by the condition $ x_+ = \lambda $ (grey dot-dashed line), as can be deduced from Fig.~\ref{fig:optimisation}e.
{All points within the regime of strong TUR violation satisfy $ 0.95 \lessapprox \langle \omega \rangle/ \omega_r <1$,} indicating proximity to resonance. Accordingly, we observe that $ x_+ > \lambda $ holds throughout this region as illustrated in Fig.~\ref{fig:optimisation}e. 

Figure~\ref{fig:optimisation}d shows that, in the regime of strong TUR violation, the current $\langle \omega \rangle$ is significantly smaller than the maximum current $\omega_t$ given by Eq.~\eqref{eq:omT}, and increases as $\Omega$ decreases towards this maximum. This increase is accompanied by a corresponding rise in $\mathcal{T}$ as $x_+$ approaches $\lambda$. Since the entropy production $\sigma = \sigma_c + \sigma_{2L}$ is proportional to the average current $\langle \omega \rangle$ (see Eqs.~\eqref{eq:sigmac} and \eqref{eq:sigma2L}), this panel also reflects the relative magnitude of $\sigma$.

Figure~\ref{fig:optimisation}f demonstrates that the increase in fluctuations observed with decreasing $\Omega$ can be attributed to a reduction in the magnitude of the current-taming force, $\partial_\alpha U$, exerted by the potential $U$. This force is estimated as $\max F_\alpha = \lambda \Omega (x_+ - \lambda)$.

The sharp, piecewise-linear boundaries observed in the region of strong TUR violation in Fig.~\ref{fig:optimisation} are a consequence of the relatively coarse sampling of the parameter space. The apparent fuzziness near the points $[\Omega,\lambda]=[1,0.8]$ and $[1.5,1.5]$ stems from finite numerical precision---attributable to the limited simulation runtime---as well as enhanced current fluctuations in these regions.

In our setting, described in Tab.~\ref{tab:parameters}, generating the data for Fig.~\ref{fig:optimisation} required more than ten days of computation, making further improvements in numerical precision beyond our current computational resources. Our simulations also suggest that the global minimum of $\mathcal{T}_\infty$ in the parameter regime of Tab.~\ref{tab:parameters} is on the order of $5\times 10^{-4}$. However, this value is below the numerical precision of approximately $10^{-3}$ achieved in our simulations.

The finding that the most stable operating regime of the engine coincides with the resonant regime, where $x_+ > \lambda$ and $\langle \omega \rangle \approx \omega_r$, is significant for practical realizations of the machine. This correspondence enables fine-tuning of the parameters using only measurements of the average current, which are much less demanding than measuring long-time fluctuations.

\section{Possible experimental implementation of the model}
\label{sec:experiment}

\begin{figure}[h!]
    \centering
\includegraphics[width=0.5\textwidth]{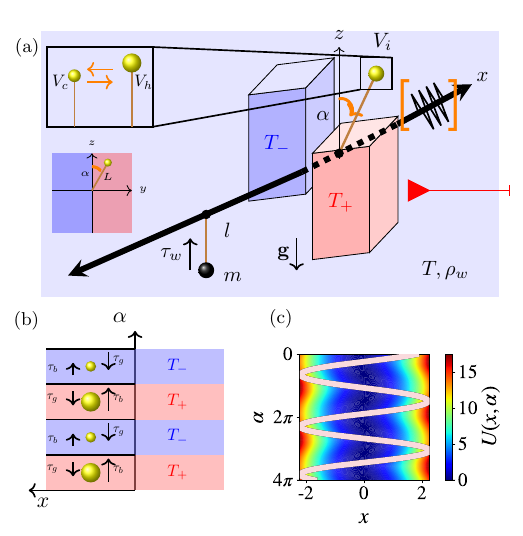}
\caption{\emph{Possible experimental realization of the considered setup.} (a) Schematic of the engine. A stochastic pendulum, parameterized by its angular coordinate \(\alpha\), is composed of a thermoresponsive microgel with mass \(M\) and is immersed in water at temperature \(T_-\). At \(T_-\), the microgel is in a collapsed state, denser than water (\(\rho_w\)), and thus sinks. When \(\alpha \in [0, \pi)\), a laser heats the pendulum to \(T_+\), triggering a transition to the swollen state, in which the gel becomes less dense than water and floats, as illustrated in the inset and in panel (b), where \(\tau_g\) and \(\tau_b\) denote the torques induced by gravity and buoyancy, respectively. The resulting buoyant force generates a net torque that lifts the weight \(m\), producing a torque \(\tau_w\) on the rod. The rotating rod is also coupled to a spring with extension \(x\), whose equilibrium position depends on \(\alpha\), generating the potential \(U\) shown in panel (c) (parameters listed in Tab.~\ref{tab:parameters}). The solid line in panel (c) represents a sample stochastic trajectory of the system, which closely follows the potential minimum.
}  
     \label{fig:MODEL_SCHEME}
\end{figure}

The model described in Sec.~\ref{sec:model} can be experimentally realized using the setup illustrated in Fig.~\ref{fig:MODEL_SCHEME}. The degree of freedom \(\alpha\) represents the angular position of a pendulum consisting of a massless rod with a mass \(M\) attached at its end. This pendulum is mounted on a solid rod and immersed in water. It is thus pulled downward by gravity and pushed upward by buoyancy, inducing a torque $\tau_d$ on the rod. Attached to the same rod is a weight with mass \(m\), which induces a torque \(\tau_w\) on the rod. The potential energy of this weight in the gravitational field acts as the work reservoir in the system.

The pendulum is driven by a heat engine mechanism operating as follows: the surrounding water is held at a constant temperature \(T_-\). The mass \(M\) consists of a thermoresponsive microgel that swells or shrinks depending on the local temperature~\cite{Zakrevskyy2012}. This behavior is governed by the transition rates given in Eqs.~\eqref{eq:k_+} and~\eqref{eq:k_-}, where the energy difference in the exponent accounts for both the intrinsic energy difference \(\Delta \epsilon\) between the two states and the work \(U_+(\alpha) - U_-(\alpha) - \Delta \epsilon\) required for expansion against water pressure at a given depth. The term \(\log G\) represents the increase in microgel entropy upon swelling. In the swollen state with volume \(V_h\), the gel has a lower density \(m/V_h\) than the water density \(\rho_w\), so the pendulum moves upward. In the collapsed state, its density \(m/V_c\) is higher than \(\rho_w\), causing the pendulum to sink. By heating the mass with a laser when \(\alpha \in [0, \pi)\) and cooling it when \(\alpha \in [\pi, 2\pi)\), and assuming rapid swelling and collapsing dynamics, a net torque is generated that rotates the pendulum and rod opposite to the torque \(\tau_w\).

This setup is  described by Eq.~\eqref{eq:om} when the potential \(U\) vanishes and \(U_d(\alpha)\) is of the form
\begin{equation}
    U_d(\alpha) = \tau_d (1 + \cos \alpha) + \delta_{d +} \Delta \epsilon.
    \label{eq:Udexp}
\end{equation}
Due to the cosine term above, the current \(j(t)\) is nonstationary and its fluctuations are stronger than in the case of the potential \(U_d(\alpha)\) in Eq.~\eqref{eq:Ud} with \(\left|\alpha - \pi\right|\) instead of \((1 + \cos \alpha)\), and thus the TUR cannot be broken in this case. We have verified this intuitive conclusion by computer simulations both in the situation with \(U = 0\) and with \(U\) given by Eq.~\eqref{eq:UPitzonka} with parameters from Tab.~\ref{tab:parameters} (data not shown and available upon request).

Hence, in order to violate the TUR, the experimental setup must be modified such that the pendulum performs a linear ``up/down'' motion rather than a rotational one, with rapid reversals at the uppermost and lowermost points of its trajectory. {Note that these reversals can in fact be smooth, provided that they occur over a length scale much shorter than the characteristic scale on which the pendulum velocity changes appreciably. Furthermore, the non-differentiable potential \(\left|\alpha - \pi\right|\) is not pathological: the discontinuity in its derivative simply reflects that gravity acts opposite to increasing $\alpha$ for $\alpha < \pi$, and along increasing $\alpha$ for $\alpha > \pi$.}

{The most difficult task would be to supplement the above-described setup with the additional `current taming' potential \(U\) in Eq.~\eqref{eq:UPitzonka}.} This could be achieved by connecting the rotating rod to a suitable spring, the equilibrium position of which would be coupled by a feedback loop to the angular position of the mass. Alternatively, this could be achieved by charging some part of the aperture or making it magnetic, and applying an electric or magnetic field with the same properties as the above-described spring. {In any case, the realization is within current experimental capabilities.}

The parameter regime considered in our numerical simulations requires fine-tuning of the swollen and collapsed volumes of the microgel so that the overall torque generated by the heat engine is constant: $\rho_w V_h - M = M - \rho_w V_c$. Usually, the swollen volume is considerably larger than the collapsed one, $V_h \gg V_c$. Under such circumstances, the condition $\rho_w(V_h - V_c) = 2M$ can be fulfilled by suitably tuning the total mass—for example, by adding to the polymer some nonswellable element of appropriate density. Another issue would be that the system needs to be large enough so that its dynamics is not overdamped.

The dimensionless representation introduced in Sec.~\ref{sec:model} can be related to the experimentally relevant setup described here by measuring length in units of $L$, time in units of $\sqrt{\dfrac{k_{B}T_-}{M L^{2}}}$, energy in units of $k_B T_-$, and temperature in units of $T_-$; these conversions can also be inverted to recover physical units.

\section{Conclusions}
\label{sec:Conclusion}

We have introduced a physically realizable model of an autonomous stationary heat engine composed of two continuous underdamped degrees of freedom coupled to a two-level system. This nonlinear setup generates a persistent steady-state current that markedly violates classical thermodynamic uncertainty relations (TURs) and circumvents the traditional trade-offs between power, efficiency, and precision across a broad parameter regime~\cite{pietzonka2018universal}. Our results show that these violations originate from resonant coupling between the underdamped modes, where one effectively acts as an internal periodic drive~\cite{HolubecMaxPower2018,Pietzonka2022}. When this coupling is removed, the system reverts to standard TUR behavior consistent with overdamped dynamics or free diffusion under constant drift~\cite{Fischer2020}.

{These findings ultimately stem from the fact that inertia can induce persistent currents and thereby break standard detailed balance. In underdamped dynamics, velocity is odd under time reversal, which fundamentally changes the symmetry properties of the dynamics compared to overdamped systems. As a consequence, the presence of inertia can lead to systematic violations of the original TUR~\cite{Vu2019}.} In this sense, inertia plays a role analogous to quantum coherence in quantum thermodynamic systems~\cite{Meier2025}, where coherence likewise breaks detailed balance and enables similar deviations from conventional bounds. However, while both mechanisms break detailed balance, the classical route via inertia is experimentally more accessible and robust against environmental decoherence, offering a realistic platform for constructing precise autonomous heat engines or clocks without requiring fragile quantum effects.

Finally, we demonstrate that the strongest suppression of current fluctuations occurs in a resonance regime that can be identified solely from mean current measurements. Since mean currents are far easier to measure than fluctuations, this provides a practical strategy for pinpointing operating conditions where TUR violations are strongest. Taken together, our results open new pathways for surpassing classical thermodynamic bounds in underdamped systems and offer a guiding principle for the design of high-performance stochastic machines—both autonomous and externally driven.

\section*{Acknowledgements} EPC and VH were supported by Charles University (project PRIMUS/22/SCI/09 and and GAUK grant 110-10/252750). We are grateful to Patrick Pietzonka for inspiring discussions, which initiated this project.

\bibliography{references}

\appendix        

\section{Intuitive insight into current stabilization through $U$}
\label{appx:intuitionU}

The effect of the fluctuation suppression induced by the potential \(U\) can be explained using the overdamped model
\begin{equation}
\gamma\dot{z} = F - k\left(z - \frac{F t}{\gamma}\right) + \sqrt{\frac{2T}{\gamma}}\, \zeta,
\label{eq:z}
\end{equation}
in which the current $z$ is stabilized at its average value $F t/\gamma$ by the time-dependent potential $k(z - Ft/\gamma)^2/2$, which `slides' at the same pace $F/\gamma$ as the particle described by the equation. The stationary current \( j(t) = \frac{z(t + t') - z(t')}{t} \), measured after a long time \( t' \) from the initialization of the process, has average \( \frac{F}{\gamma} \) and variance 
\(\mathrm{Var}[j(t)] = \langle \tilde{z}^2(t) \rangle + \langle \tilde{z}^2(t') \rangle + 2 \langle \tilde{z}(t) \tilde{z}(t') \rangle = \frac{2T}{k t^2} \), where \(\tilde{z} = z - \frac{F t}{\gamma}\) is a stationary Ornstein-Uhlenbeck process with variance \( \frac{T}{k} \). We neglected the cross-correlation term \( \langle \tilde{z}(t) \tilde{z}(t') \rangle \) since it exponentially vanishes with increasing measurement time \( t \).
 The left-hand side $\mathrm{Var}[j(t)]/\langle j(t) \rangle^2$ of the TUR~\eqref{eq:TUR} thus decays with measurement time as $1/t^2$. On the other hand, the average stationary entropy production in this system reads $\sigma = \langle (\gamma \dot{z} - \sqrt{2T/\gamma}\, \zeta)\, \dot{z} \rangle = F^2/(\gamma T)$. Hence, the right-hand side $2/(\sigma t)$ of the TUR decays only as $1/t$. The TUR can be rearranged as $\mathrm{Var}[j(t)] \sigma/(2\langle j(t) \rangle^2) = 1/(kt) \ge 1$ and is thus invalid for measurement times $t \ge 1/k$.

\section{Current variance}
\label{appx:j2}

Here, we provide for the sake of self-consistency, the derivation of the variance of the stationary time-averaged current $j(t) = \frac{1}{t}\int_0^t dt' \omega(t')$ for the simplified version of the process in Eqs.~\eqref{eq:om}--\eqref{eq:T}, described by the Langevin equation
\begin{equation}
\dot{\omega} =  \tau_0 - \tau_{w} - \gamma_{\alpha} \omega + \sqrt{2D_{\alpha}\gamma_{\alpha}^2}\,\xi_\alpha .
\end{equation}

The stationary variance of $j(t)$ equals the stationary variance of current $\tilde{j}(t) = \frac{1}{t}\int_0^t dt' \tilde{\omega}(t')$ for the centered process \(\tilde{\omega}(t) = \tilde{\omega}(t) - \langle\omega\rangle = \omega(t) - \frac{\tau_0 - \tau_w}{\gamma_{\alpha}}\), which satisfies
\begin{equation}
    d\tilde{\omega}_t = -\gamma_{\alpha} \tilde{\omega}_t\,dt + \sigma\,dW_t.
\end{equation}
The corresponding stationary autocovariance of is
\begin{equation}
    \langle \tilde{\omega}(s) \tilde{\omega}(u) \rangle = D_{\alpha} \gamma_{\alpha} e^{-\gamma_{\alpha}|s - u|}.
\end{equation}
The current variance then reads~\cite{Fischer2020}
\begin{multline}
    \mathrm{Var}[j(t)] = \left\langle \left( \frac{1}{t} \int_0^t \tilde{\omega}(s)\,ds \right)^2 \right\rangle 
    = \frac{1}{t^2} \int_0^t \int_0^t \langle \tilde{\omega}(s) \tilde{\omega}(u) \rangle\,ds\,du \\
    = \frac{D_{\alpha} \gamma_{\alpha}}{t^2} \int_0^t \int_0^t e^{-\gamma_{\alpha}|s - u|} ds\,du
    = \frac{2 D_{\alpha}}{t} \left( 1 + \frac{e^{-\gamma_{\alpha} t} - 1}{\gamma_{\alpha}t} \right).
    \label{eq:varjappx}
\end{multline}

\section{Dynamics of $x$ and $v$}
\label{appx:xv}

In this section, we derive approximate analytical formulas for probability densities for position $x$ and velocity $v$ shown in Fig.~\ref{fig:2Dseries}.

Stationary solution to Eq.~\eqref{eq:v} for vanishing noise ($D_x = 0$) and $\alpha(t) = \langle \omega \rangle t$ is
\begin{equation}
    x(t) = x_0 \left[s_+  \sin(\Omega  \langle \omega \rangle  t) - c_+  \cos(\Omega  \langle \omega \rangle  t)\right],
    \label{eq:xt}
\end{equation}
where $s_+ = \kappa_{\alpha x} + \kappa_x - \Omega^2  \langle \omega \rangle^2$, $c_+ = \gamma_x    \Omega  \langle \omega \rangle$, and $x_0 = \kappa_{\alpha x}\lambda/(c_+^2 + s^2)$.
The corresponding velocity, $v(t) = \dot{x}(t)$, and acceleration, $a(t) = \ddot{x}(t)$, are 
\begin{align}
v(t) &= x_0\Omega  \langle \omega \rangle \left[c_+  \sin(\Omega  \langle \omega \rangle  t) + s_+  \cos(\Omega  \langle \omega \rangle  t)\right],
\label{eq:vt}\\
a(t) &= x_0\Omega^2  \langle \omega\rangle^2  \left[ c_+  \cos(\Omega  \langle \omega \rangle  t) -s_+  \sin(\Omega  \langle \omega \rangle  t)\right].
\end{align}

The probability densities for the above position $x$ and velocity $v$ are given by
\begin{align}
P(x) &= N_x \frac{1}{v[x]},
\label{eq:px}\\
P(v) &= N_v \frac{1}{a[v]},
\label{eq:pv}
\end{align}
where $N_x$ and $N_v$ are normalization constants, $v(x) \equiv v\left[ t(x)\right]$, and  $a(x) \equiv a\left[ t(v)\right]$. 

The supports $x \in [-x_+, x_+]$ and $v\in [-v_+,v_+]$ of the two distributions are given by amplitudes $x_+$ and $v_+$ of the position and velocity waves in Eqs.~\eqref{eq:xt} and \eqref{eq:vt}: 
\begin{align}
x_+ &= \frac{\kappa_{\alpha x} \lambda}{\sqrt{s_+^2+c_+^2}},
\label{eq:xp}\\
v_+ &= x_+ \Omega  \langle \omega \rangle.
\label{eq:vp}
\end{align}
Maximum $|x(t)|$ is attained when $v(t) = 0$ and maximum $|v(t)|$ when $a(t) = 0$. The oscillations exhibit a resonance with maximum amplitude \(x_+\) when the average angular velocity \(\langle \omega \rangle\) equals
\begin{equation}
    \omega_r = \frac{\sqrt{\kappa_{\alpha x} + \kappa_x - \gamma_x^2 / 2}}{\Omega}.
    \label{eq:omR}
\end{equation}

To account for the noise in
Eq.~\eqref{eq:v}, we calculated the probability densities $P(x)$ and $P(v)$ shown Fig.~\ref{fig:2Dseries} as convolutions of $p(x)$ and $p(v)$ with Gaussian Kernels with effective temperatures $T_x$ and $T_v$, respectively. To be specific,
\begin{equation}
P(y) = \frac{1}{\sqrt{2\pi N_y}}
\int_{y_-}^{y_+} p(y') \exp{\left(- \frac{(y-y')^2}{2 T_y} \right)} dy',
\label{eq:PConv}
\end{equation}
where $y = x$ or $v$. A good agreement with the data was obtained for $T_x = 1/600 = 1/(\gamma_x + \gamma_{\alpha x})$ and $T_v = 1 = T_-$. While we do not have intuition why the effective temperature for $x$ is close to $1/(\gamma_x + \gamma_{\alpha x})$, $T_v = T_-$ is the actual noise temperature in Eq.~\eqref{eq:v}.

\end{document}